\documentclass[sigconf]{acmart}

\usepackage{graphicx}
\usepackage{tabularx}
\renewcommand\footnotetextcopyrightpermission[1]{}
\AtBeginDocument{%
  \providecommand\BibTeX{{%
    \normalfont B\kern-0.5em{\scshape i\kern-0.25em b}\kern-0.8em\TeX}}}

\setcopyright{acmcopyright}
\copyrightyear{2018}
\acmYear{2018}
\acmDOI{10.1145/1122445.1122456}

\acmConference[HRI ’21 Companion]{Companion of the 2021 ACM/IEEE International Conference on Human-Robot Interaction (HRI ’21 Companion)}{March 08--11, 2021}{virtual}
\acmBooktitle{Companion of the 2021 ACM/IEEE International Conference on Human-Robot Interaction (HRI ’21 Companion),
  March 08--11, 2021, virtual}
\acmPrice{15.00}
\acmISBN{978-1-4503-XXXX-X/18/06}

\begin{document}

\title{Personalized Affect-Aware Socially Assistive Robot Tutors Aimed at Fostering Social Grit in Children with Autism}

\author{Zhonghao Shi}
\email{zhonghas@usc.edu}
\orcid{https://orcid.org/0000-0002-2855-5863}
\affiliation{%
  \institution{University of Southern California}
  \city{Los Angeles}
  \state{USA}
  \postcode{90007}
}

\author{Manwei Cao}
\email{manweica@usc.edu}
\orcid{}
\affiliation{%
  \institution{University of Southern California}
  \city{Los Angeles}
  \state{USA}
  \postcode{90007}
}

\author{Sophia Pei}
\email{sophiakp@usc.edu}
\orcid{}
\affiliation{%
  \institution{University of Southern California}
  \city{Los Angeles}
  \state{USA}
  \postcode{90007}
}

\author{Xiaoyang Qiao}
\email{xiaoyanq@usc.edu}
\orcid{}
\affiliation{%
  \institution{University of Southern California}
  \city{Los Angeles}
  \state{USA}
  \postcode{90007}
}

\author{Thomas R Groechel}
\email{groechel@usc.edu}
\orcid{}
\affiliation{%
  \institution{University of Southern California}
  \city{Los Angeles}
  \state{USA}
  \postcode{90007}
}

\author{Maja J Matari\'c}
\email{mataric@usc.edu}
\orcid{https://orcid.org/0000-0001-8958-6666}
\affiliation{%
  \institution{University of Southern California}
  \city{Los Angeles}
  \state{USA}
  \postcode{90007}
}
\renewcommand{\shortauthors}{Trovato and Tobin, et al.}

\begin{abstract}
Affect-aware socially assistive robotics (SAR) tutors have great potential to augment and democratize professional therapeutic interventions for children with autism spectrum disorders (ASD) from different socioeconomic backgrounds. However, the majority of research on SAR for ASD has been on teaching cognitive and/or social skills, not on addressing users' emotional needs for real-world social situations. To bridge that gap, this work aims to develop personalized affect-aware SAR tutors to help alleviate social anxiety and foster {\it social grit}--the growth mindset for social skill development--in children with ASD. We propose a novel paradigm to incorporate clinically validated Acceptance and Commitment Training (ACT) with personalized SAR interventions. This work paves the way toward developing personalized affect-aware SAR interventions to support the unique and diverse socio-emotional needs and challenges of children with ASD.
\end{abstract}

\begin{CCSXML}
<ccs2012>
<concept>
<concept_id>10010520.10010553.10010554</concept_id>
<concept_desc>Computer systems organization~Robotics</concept_desc>
<concept_significance>500</concept_significance>
</concept>
</ccs2012>
\end{CCSXML}

\ccsdesc[500]{Computer systems organization~Robotics}

\keywords{Human-Robot Interaction, Socially Assistive Robotics, Affective Computing}

\maketitle
\pagestyle{plain}
\section{INTRODUCTION}

Socially assistive robot (SAR) learning companions and tutors have great potential to effectively promote cognitive development of social skills for children with ASD~\cite{saleh2020robot}. Past research on SAR for ASD has been mainly focusing on augmenting or automating interventions for specific cognitive impairments of ASD, such as social interaction deficits and stereotyped behaviors~\cite{ismail2019leveraging}. This further validates the potential of autonomous SAR to provide more cost-effective and accessible clinical treatments for children with ASD, so the economic burden of their families can be alleviated \cite{lavelle2014economic}. However, individuals with ASD also face significant emotional challenges such as social anxiety due to deficits in social-emotional reciprocity due to ASD~\cite{bellini2006development}. While cognitive impairments can be reduced with evidence-based persistent clinical interventions and personal practice~\cite{white2007social}, \citet{sumiya2018emotions} showed that children with ASD tend to have low social motivation and grit, making it especially challenging for them to persistently practice social skills in daily social interactions, despite having previously learned and practiced those skills with therapists~\cite{briot2020social}. As shown in a study conducted by~\citet{bellini2006development}, children with ASD are at high risk of suffering social anxiety, so they can particularly demonstrate this lack of generalization of learned social skills into their daily social life. SAR therefore also has the potential to provide support for addressing the socio-emotional challenges faced by children with ASD.

In this work, we propose a novel paradigm that incorporates clinically validated Acceptance and Commitment Training (ACT) methods~\cite{eifert2005acceptance} with personalized SAR interventions to help children with ASD alleviate socio-emotional challenges and foster social grit, so they can become more emotionally competent and more likely to positively engage in social activities.

\section{Methods}
% \subsection{ACT-Based SAR Interventions for social anxiety and grit}
\label{sec:act}

\begin{figure*}[t!] 
    \centering
    \includegraphics[width=1\linewidth]{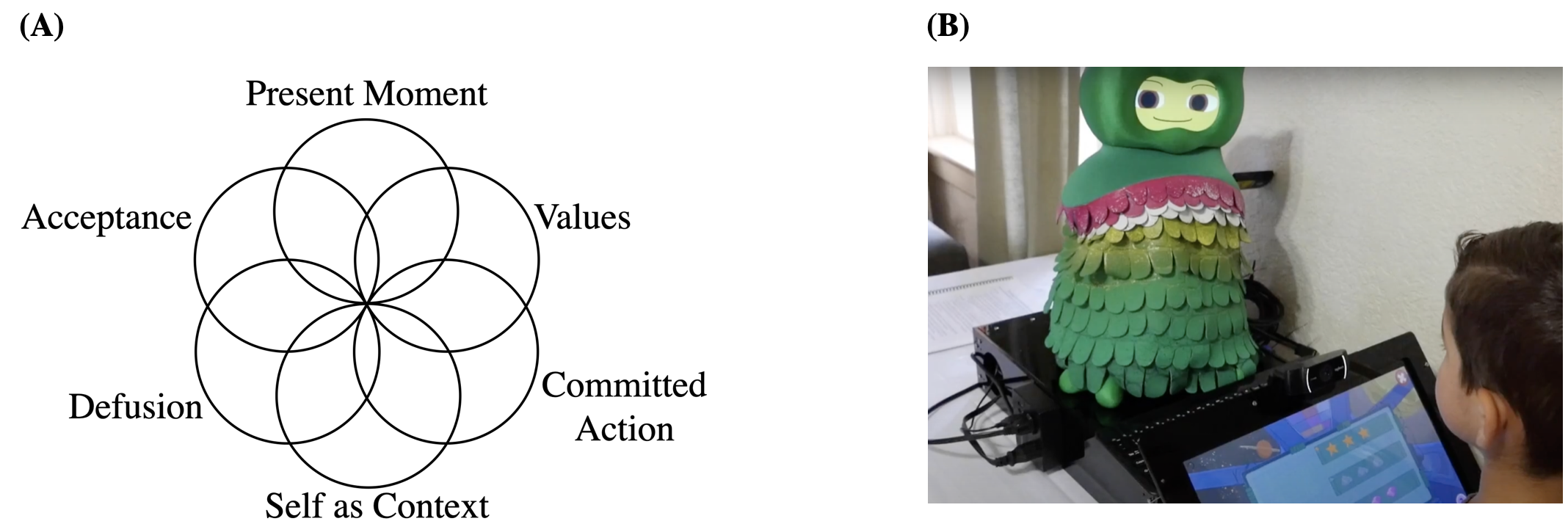}
    \vspace{1.5em}
\caption{\textbf{(A)} Six areas of focus of ACT methods~\cite{hayes2009acceptance}; \textbf{(B)} Our autonomous SAR system for children with ASD} 
    \vspace{1.5em}
    \label{fig:system}
\end{figure*}

% A: Based on Kort et al. [33], two-dimensional arousal and valence framework for cognitive-affectivestates. States with positive valence (more pleasurable) are on the right; states with negative valence (moreunpleasant) are on the left. Similarly, states with positive arousal (more constructive learning activity) areat the top; states with negative arousal (no learning activity) are at the bottom.B: Ideal circular flow ofcognitive-affective states

\begin{table*}[t!]
\centering 

\begin{tabularx}{\textwidth}{|p{3.3cm}|p{5cm}|X|}
\hline
\textbf{ACT Areas of Focus\cite{eifert2005acceptance}} & \textbf{ACT Design Principles~\cite{eifert2005acceptance}} & \textbf{ACT-Based SAR Interventions~\cite{eifert2005acceptance}} \\
\hline
Acceptance & Increasing the participants' social behaviors replacing avoidance/escape behavior in the presence of aversive social stimuli in order to contact a desired reinforcer. & A SAR intervention is designed to walk the participant through the “Tug-of-war with an Anxiety Monster” metaphor~\cite{eifert2005acceptance}. The SAR dialog is scripted and programmed to help uer user fully embrace their aversive thoughts and feelings in the presence of aversive social stimuli in a non-judgemental and empowering way. \\
\hline
Defusion & Weakening the participants' rigid rule-following behaviors and restrictive cognitive patterns, and increasing a skill repertoire of flexible responding to social stimuli that were previously rigidly responded to. & A SAR intervention is designed to engage the user in the "Hand" exercise~\cite{eifert2005acceptance} and "Silly Voices" exercise~\cite{eilers2015exposure}. The goal of the SAR tutor is to help the user foster their ability to create space and distance between themselves and their negative emotions. \\
\hline
Present moment & Strengthening the user's skill repertoire of actively attending to social stimuli in the present moment, while weakening the skill repertoire of attending to thir own emotions/memories/thoughts with respect to past/future/imagined events. & A SAR intervention is designed to guide the user through the "Soles of the Feet" mindfulness exercise~\cite{singh2003soles}, where the SAR tutor teaches the user about how to shift attention flexibly to focus on the present moment, instead of overthinking about the negative past experience or uncertainty of the future. \\
\hline
Self-as-context & Weakening the user's rigid concepts regarding self image and self identities, and increasing their flexible perspective-taking skills of relational responding to I/you, here/there, now/then. & A SAR intervention is designed to guide the user through the "Bus Driver" exercise~\cite{eifert2005acceptance}. In this exercise, the SAR tutor's goal is help the user understand that their actions are defined by their values rather than by the external environment. \\
\hline
Values & Enhancing the user's skill repertoire of deriving and following rules that associate previously aversive stimuli with a value/motivation, or increasing effectiveness of existing value/motivation. & A SAR intervention is designed to guide the user through the "Three Wishes" exercise~\cite{eifert2005acceptance} with the goal of increasing the user's ability to identify their intrinsic motivation and life values as the reason to engage in meaningful social activities. \\
\hline
Committed Action & Increasing the user's response repertoire of meaningful social behaviors that are in line with their values. & A SAR intervention is designed to guide the user through the "Choice Point" exercise~\cite{bailey2014weight}, so that the user is likely to perform actions that move them closer toward their values and goals, and away from experiential avoidance. \\
\hline
\end{tabularx}

\caption{Example SAR interventions based on the design principles of ACT~\cite{niles2014cognitive, eifert2005acceptance, hayes2020liberated}} 
\label{tab:interventions}
\end{table*}

We propose a novel paradigm to design personalized SAR interventions based on clinically validated acceptance and commitment training (ACT) methods~\cite{eifert2005acceptance} to help children with ASD with their socio-emotional challenges and lack of social grit. ACT is a behavioral treatment that has been clinically validated and is commonly conducted in clinical settings for the treatment of socio-emotional challenges~\cite{dalrymple2007acceptance, niles2014cognitive}. ACT aims to promote mindfulness, acceptance, and cognitive defusion (learning to detach from current thoughts and emotions) with the ultimate goal of increasing psychological flexibility and promoting behavior change that aligns with one’s life values~\cite{hayes2009acceptance}.  Our goal of incorporating ACT with SAR interventions is to not only support the socio-emotional needs of children with ASD but also to foster social grit and encourage behavior changes toward more actively engaging in positive social activities.

Specifically, we adapt the ACT treatment format outlined in a prior clinical study by \citet{niles2014cognitive} and a clinical manual by~\citet{eifert2005acceptance} to inform the design of our personalized SAR interventions. As detailed in Figure~\ref{fig:system}(A) and Table~\ref{tab:interventions}, by following the design principle of the six areas of focus in ACT, we first select ACT exercises that can be potentially effectively adapted and performed by a SAR tutor, and then design SAR interactions based on those interventions. Our design involves weekly intervention sessions wherein the child engages in a triadic interaction with the robot and caregiver to complete ACT-based SAR interventions. The affect-aware SAR tutor is programmed to conduct interventions while providing personalized feedback to keep the participants engaged and motivated in the exercises. The caregiver helps the child when the child's needs are outside of the scope of the SAR's functionalities. More details about the SAR system and experiment design are found in the next section.

\section{Experiment and System Setup}
\label{sec:setup}
\subsection{Study Design}
The goal of this project is to study how effectively personalized affect-aware SAR can deliver ACT-based interventions to children with ASD outside clinical settings. We will evaluate the effectiveness of the SAR-supported interventions to help alleviate social anxiety and foster social grit for children with ASD. Our study design follows the single-subject withdrawal design embedded in a multiple-baseline design commonly used in ASD studies~\cite{cardon2011deciphering}. The study consists of three phases: pre-treatment (3-5 weeks), in-treatment (6-10 weeks), and post-treatment (2-4 weeks). Each participant stays in the pre-treatment (baseline) phase for 3 weeks or longer and then enters the in-treatment phase at different points of time. Consequently, a comparison measurement between  participants'  baseline level of performance is established, and the potential maturation effect caused by passage of time can be controlled~\cite{cardon2011deciphering}. The in-treatment phase involves the deployments of ACT-based personalized SAR interventions. We plan to conduct weekly sessions to cover the six areas of focus in ACT treatment summarized in Sec~\ref{sec:act}. Since the socio-emotional needs of each child are unique, the types and total number of treatment sessions are personalized based on the child's behavioral improvements during the in-treatment phase. The in-treatment phase can be extended with more sessions if the SAR decision-tree-based personalization algorithm, co-designed by the therapist and researcher, determines the child needs more support to reach their therapeutic goal. The final post-treatment phase allows the participants' behavior changes to be re-assessed and studied to learn whether observed improvements can be sustained after the removal of the intervention.

\subsection{Personalized SAR Intervention System}

As shown in Figure~\ref{fig:system}(B), we plan to develop the SAR tutor with a touchscreen tablet that will serve as a shared collaborative space for the child and robot. Each participant will take part in a first pilot session, whose multimodal data will be annotated and used to pre-train personalized affective models of arousal and valence using supervised domain adaptation we have already validated in our previous work~\cite{shi2021toward}. Using real-time predictions of arousal and valence, we plan to apply an affective reinforcement learning approach in order to personalize the robot's responses to keep the child engaged in the ACT interventions. The autonomous SAR system will use dialog to further encourage social grit and a growth mindset grounded in prior work~\cite{park2017growing, rhew2017effects} in order to keep the participants motivated throughout the intervention.  

\subsection{Participant Information}
In this study, we plan to recruit 14 participants with the following inclusion criteria: 1) age between 7 and 12 years old; 2) good physical, sensory (hearing, vision), and medical health conditions; 3) English as the primary language; 4) clinical diagnosis of ASD as described in the Diagnostic and Statistical Manual of Mental Disorders--Version 5 \citep{van1992comparison,kanne2008diagnostic, dover2007diagnose, baird2003diagnosis}; 5) at high risk of suffering social anxiety based on the Anxiety Disorders Interview Schedule for DSM-IV (ADIS-IV)~\cite{grisham2004anxiety}; 6) at high risk for low social grit based on the assessment of acceptance and action questionnaires~\cite{hayes2006acceptance}. The detailed study plan including the recruitment procedure, consent process, participant privacy and data confidentiality protocol is currently being developed, and will be submitted to Institutional Review Board of the University of Southern California for review.

\subsection{Assessment}
To assess the participants' behavior changes during the in-treatment phase, we plan to administer two validated assessment tools: 1) Acceptance and Action Questionnaire (16-item version) developed specifically to measure the proposed mechanisms of change in ACT treatment~\cite{hayes2004measuring}; and 2) Self-Statements During Public Speaking Questionnaire (SSPS) developed to measure negative and positive self-statements in the context of public speaking in social situations~\cite{hofmann2000instrument}. These measurements also inform the personalization of the SAR tutor, so the duration and design of in-treatment phase can be personalized to each individual child's needs.

To validate the effectiveness of the personalized SAR intervention for emotional challenges and social grit, we plan to compare the participants' outcome measures between the pre-treatment and post-treatment phases. We will use three widely used validated outcome assessment tools for social anxiety: 1) the Liebowitz Social Anxiety Scale–Self-Report (LSAS-SR): a 24-item measure that assesses fear and avoidance of social interaction and performance situations~\cite{fresco2001liebowitz}; 2) the Social Interaction Anxiety Scale (SIAS): a 20-item measure of cognitive, affective, or behavioral reactions to social interaction in dyads or groups~\cite{mattick1998development}; and 3) the Social Phobia Scale (SPS): a 20-item measure describing situations or themes related to being observed by others~\cite{mattick1998development}. Moreover, to study the effects of SAR interventions on social grit and motivation, we also plan to conduct direct measurements of the frequency of the social activities engaged by the participants during the study. Semi-structured interviews with participants' family members (parents, teachers, friends) will also be conducted weekly to collect  indirect measures. Statistical analysis with multilevel modeling \cite{kenny1998handbook, niles2014cognitive} will then be conducted to evaluate the significance of behavior changes between the pre-treatment and post-treatment phases.

\section{ACKNOWLEDGMENTS}
This work is supported by the NSF Expeditions in Computing grant on Socially Assistive Robotics, IIS-1139148.

\bibliographystyle{ACM-Reference-Format}
\bibliography{sample-base}
\end{document}